\documentclass[twoside]{ilcws07}
\usepackage[latin1]{inputenc}
\usepackage[dvips]{graphicx,epsfig,color}
\usepackage{wrapfig,rotating}
\usepackage{amssymb,amsmath,array}

\begin{document}
\title{
High Resolution Cavity BPM for ILC Final Focal System (IP-BPM)} 
\author{Tomoya Nakamura$^1$, Yosuke Honda$^2$, Yoichi Inoue$^3$, Toshiaki Tauchi$^4$, Junji Urakawa$^2$,\\Tomoyuki Sanuki$^5$, Sachio Komamiya$^1$ 
\vspace{.3cm}\\
1- The University of Tokyo - Dept of Physics \\
7-3-1, Hongo, Bunkyo-ku, Tokyo - Japan
\vspace{.1cm}\\
2- KEK - ATF \\
1-1 Oho, Tsukuba, Ibaraki - Japan
\vspace{.1cm}\\
3- Tohoku Gakuin University - Dept of Engineering\\
1-3-1 Tsuchitoi, Aoba-ku, Sendai, Miyagi - Japan
\vspace{.1cm}\\
4- KEK - Institute of Particles and Nuclear Studies \\
1-1 Oho, Tsukuba, Ibaraki - Japan
\vspace{.1cm}\\
5- Tohoku University - Dept of Physics \\
6-3 Aoba, Aramaki, Aoba-ku, Sendai, Miyagi - Japan\\
}


\maketitle

\begin{abstract}
IP-BPM (Interaction Point Beam Position Monitor) is an ultra high resolution cavity BPM to be used at ATF2, a test facility for ILC final focus system.
Control of beam position in 2 nm precision is required for ATF2.
Beam tests at ATF extraction line proved a 8.7 nm position resolution.
\end{abstract}

\begin{wrapfigure}{r}{0.5\columnwidth}
\centerline{\includegraphics[width=0.45\columnwidth]{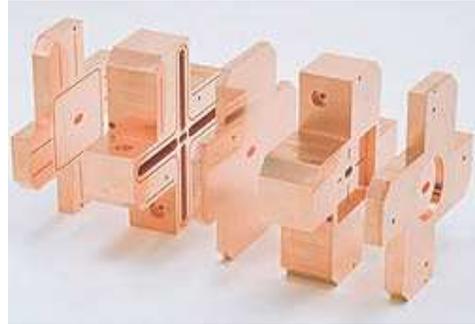}}
\caption{IP-BPM Block}\label{Fig:IPBPM}
\end{wrapfigure}

\section{ATF2}

ATF2 is an extension of ATF, a test facility for ILC accelerator development. 
ATF is the only facility today which can achieve beam emittance of the ILC specification.
Experimental studies for ILC final focus is planned at ATF2, and is starting its operation at October, 2008.

There are two main goals for ATF2.
First, achievement of 35 nm beam size at the IP.
Second, control of beam position in 2 nm precision at the IP.
Old model cavity BPMs at ATF have achieved 17 nm position resolution so far \cite{17nm}.
But to achieve the two goals, a BPM of 2 nm position resolution (IP-BPM) is required at ATF2.
We have fabricated two hot models of IP-BPM, and tested its performance at ATF.


\section{Characteristics of IP-BPM}
IP-BPM (See Figure \ref{Fig:IPBPM}) is a cavity BPM and uses di-pole mode signals to detect the beam position.
The di-pole mode signal magnetically couples to the wave guides through the slots, and the signal is read out from coaxial antennas.

IP-BPM has 3 main characteristics.
(1) To measure the beam position perfectly independent in X and Y, the cavity is designed to be rectangular. 
The designed resonant frequency of X di-pole mode is 5.712 GHz, while Y di-pole mode is 6.426 GHz.
The X-Y isolation was confirmed to be better than $-50$ dB.
(2) It is designed to have low angle sensitivity, since the angle jitter would be large at IP, due to the strong focusing.
We achieved this by designing the cavity length short (L = 6 mm).
(3) It has a high coupling to achieve ultra high position sensitivity.
The aperture of the beam pipe is designed to be small, to recover the coupling and position sensitivity which also reduces due to the short cavity length.
The design value of coupling constant $\beta$ for X, Y is 1.4 and 2.0, respectively.

\begin{figure}
\centerline{\includegraphics[width=0.8\columnwidth]{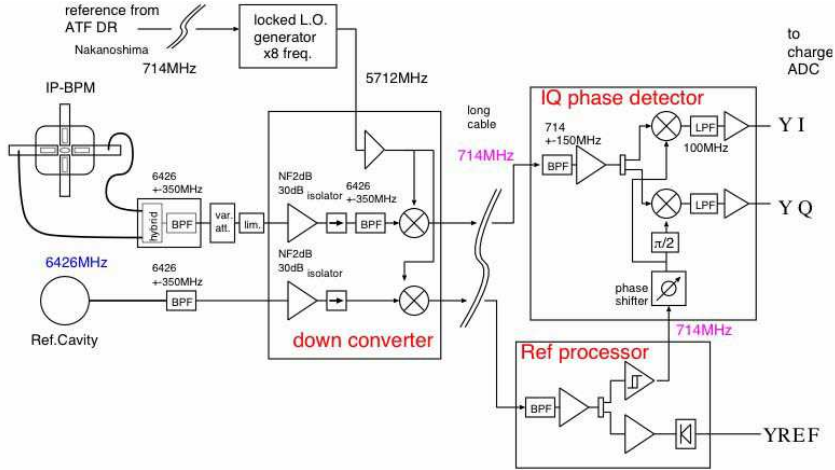}}
\caption{Measurement Scheme}\label{Fig:Detection}
\end{figure}

\section{Measurement Scheme}

As shown in Figure \ref{Fig:Detection}, IP-BPM has a sensor cavity and a reference cavity for X and Y.
The sensor cavity monitors the beam position using di-pole mode, while the reference cavity monitors the beam charge using mono-pole mode of same resonant frequency.
The RF signals (6.426 or 5.712 GHz) from the cavities are quickly down converted to 714 MHz by a local oscillator, to minimize signal loss.
The sensor signal enters a variable attenuator before the down converter, to enlarge dynamic range of the electricity, which is necessary when making calibration.
Finally, the sensor signals are detected by the phase detector.


We can acquire I signals and Q signals, which are 90 degrees different in phase.
Since beam angle signals or tilt signals of beam bunches are 90 degrees different in phase from position signals, through precise tuning we can divide position signals to I signals,
and other signals to Q signals.
To achieve this I-Q tuning, we need a beam synchronized phase origin.
The reference cavity is used for this purpose, and it is down converted by the same local oscillator as used for the sensor signal, to maintain the phase relativity between reference signal and sensor signal.
Also, to prevent the contamination of modes other than the di-pole modes, we use band-pass filters to select the di-pole mode of our concern. 

\section{Basic Tests}

2 blocks (4 cavities) have been fabricated.
Their basic performance was checked by the following experiments.
(1) Their resonant frequencies, Q vales, X-Y isolations were checked by using a network analyzer.
Also, we tuned the reference cavity frequency to match that of the average of the 3 sensor cavities.
(2) We made an R/Q measurement to estimate the cavity geometry and confirmed that the di-pole mode is sensitive to the beam position.
R/Q of the cavity was measured through a bead perturbation measurement.
\begin{wrapfigure}{r}{0.5\columnwidth}
\centerline{\includegraphics[width=0.45\columnwidth]{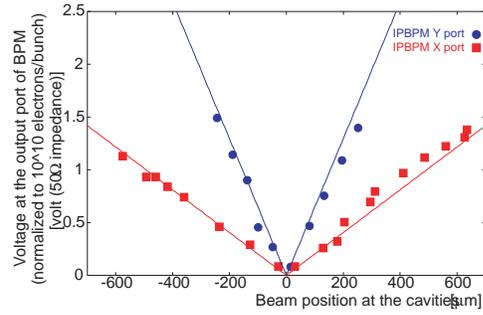}}
\caption{Position Sensitivity}\label{Fig:Vslope}
\end{wrapfigure}
(3) We carried out position and angle sensitivity tests at ATF extraction line.
At position sensitivity measurements, we swept the beam against the cavity by controlling the steering magnets.
We used a diode to detect signals, so the signal response to the beam position forms a V-shape (See Figure \ref{Fig:Vslope}).
At angle sensitivity measurements, we tilted the cavity against the beam by sandwiching shims between the BPM blocks and the stage.
Even when the beam passes the cavity center, the signal is non-zero in this situation.
This signal was compared with the equivalent position signal.
As a result, IP-BPM position sensitivity was proved to be coincident with the expected value, also shown with lines in Figure \ref{Fig:Vslope}.
Also, angle sensitivity was proved to be reduced enough not to ruin the measurement.

\section{I-Q Tuning}
 
Contamination of angle signal would degrade the position resolution greatly.
Precise decoupling of position signal and angle signal is critical for achieving ultra high position resolution.

We used 2 steering magnets to sweep the beam parallel.
From the signal response, we were able to know the relative position of the 3 sensor cavities. 
We succeeded to align 3 cavities at same height in precision of a few microns, using shims to control the cavity height.
This allowed us to keep the angle signal very small and tune the I-Q decoupling precisely.

\section{Position Resolution Measurement}
We used 3 cavities to determine the position resolution.
From the upstream, we call them BPM1, BPM2, and BPM3.
Our definition of "position resolution" is, (RMS of the residual between measured and predicted beam position at BPM2) $\times$ (Geometry Factor).
The prediction is made by using the beam information from BPM1 and BPM3.

Two types of measurements, calibration run and resolution run, were carried out.
At calibration run we swept the beam against the cavities, while at resolution run we fixed the beam position and took statistics.
Calibration run is for calibrating the I signal to the actual beam position.
In order to enlarge the dynamic range of the detecting electronics, we set the variable attenuator at 40 dB, 30 dB, and 20 dB.
We extrapolated the calibration slope for the non-attenuation case from those data.
Then, resolution run was carried out until enough statistics was achieved, especially for the non-attenuation case, 1 hour.
To determine the position resolution, we used a linear regression analysis written below:
\begin{eqnarray}
Y2I_{predicted} &=& a_0 + a_1\ast Y1I + a_2\ast Y1Q + a_3\ast Y3I + a_4\ast Y3Q + a_5\ast YREF\nonumber \\
&+&  a_6\ast X1I + a_7\ast X1Q + a_8\ast X3I + a_9\ast X3Q + a_{10}\ast XREF \nonumber \\
Residual &=& Y2I_{measured} - Y2I_{predicted} \nonumber 
\end{eqnarray}
while Y(X)iI(Q) (i=1,2,3) stands for I(Q) signals from BPMi, and Y(X)REF stands for reference signals.
The resolution will be calculated by the function below:
\begin{eqnarray}
{\rm Position\: Resolution} = {\rm Geometry\: Factor} \times \frac{{\rm RMS\: of \: Residual (ADC\: ch)}}{{\rm Calibration\: Slope (ADC\: ch/nm)}} \nonumber
\end{eqnarray}
Before the measurement, we calibrated the reference signal to beam charge.
\begin{wrapfigure}{r}{0.5\columnwidth}
\centerline{\includegraphics[width=0.45\columnwidth]{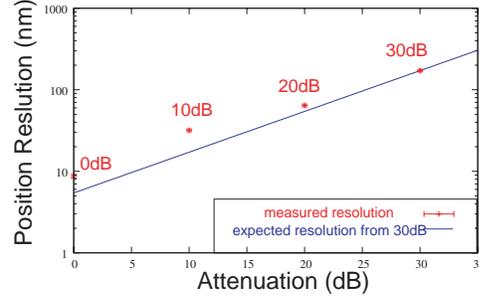}}
\caption{Position Resolution}\label{Fig:Resolution}
\end{wrapfigure}
We made an appropriate data cut of 0.640 $<$ ICT ($\times$1.6nC) $<$ 0.755.
As a result, the position resolution at non-attenuation case was proved to be 8.72 $\pm$ 0.28 $\pm$ 0.35 nm, which is the best record in the world today (See Figure \ref{Fig:Resolution}).
The beam condition was 0.68 $\times$ 10$^{10}$ e$^-$/bunch, and the dynamic range was 4.96 $\mu$m. 
This result implies that the position resolution would be 5.94 nm for the ATF2 condition (10$^{10}$ e$^-$/bunch).


In advance, thermal noise of the electricity was also checked.
Signal from the same sensor cavity was divided into 2 and detected by the same detecting scheme.
From the correlation of the two signals, we estimated the thermal noise, which determine the detecting limit of position signal.
As a result, it was estimated to be 2.57 nm, under the condition of ATF2 (10$^{10}$ e$^-$/bunch).
This result implies that by removing other noise, we can achieve resolution better than 3 nm by IP-BPM.
 
\section{Summary}
We developed a cavity BPM for the IP of ATF2, a test accelerator for ILC final focus system.
As a result of beam tests, an ultra high position resolution of 8.72 nm was proved.
Our goal is to improve the resolution to 2 nm, for nano-meter beam control at ATF2.

\section{Acknowledgements}
This work was supported by "Grant-In-Aid for Creative Scientific 
Research of JSPS (KAKENHI 17GS0210)" .
We thank for everyone involved in the ATF project for their cooperation.


\begin{footnotesize}




\begin{thebibliography}{99}
\bibitem{url} Slides: \\ 
\verb$http://ilcagenda.linearcollider.org/contributionDisplay.py?contribId=129&sessionId=97&amp;confId=1296$
\bibitem{17nm} Y. Inoue, in preparation.

\end{thebibliography}
%

\end{footnotesize}


\end{document}